\documentclass[proceedings]{JHEP37}
\usepackage{eurosym}
\usepackage{amssymb}
\usepackage{amsfonts}
\usepackage{amsmath}
\usepackage{epsfig}

\newbox\mybox

\newcommand\fverb{\setbox\mybox=\hbox\bgroup\verb}
\newcommand\fverbdo{\egroup\medskip\noindent\fbox{\unhbox\mybox}\ }
\newcommand\fverbit{\egroup\item[\fbox{\unhbox\mybox}]}
\conference{Time-dependent massless Dirac fermions in graphene}
\abstract{Using the Lewis-Riesenfeld method of invariants we construct explicit analytical solutions for the massless Dirac equation in 2+1 dimensions describing quasi-particles in graphene. The Hamiltonian of the system considered contains some explicit time-dependence
in addition to one resulting from being minimally coupled to a time-dependent vector potential. The eigenvalue equations for the two spinor components of the Lewis-Riesenfeld invariant are found to decouple into a pair of supersymmetric invariants in a similar fashion as the known decoupling for the time-independent Dirac Hamiltonians.}

\title{Time-dependent massless Dirac fermions in graphene}
\author{Boubakeur Khantoul$^{\bullet , \circ}$ and Andreas Fring$^\bullet$ \\
$^\bullet$ Department of Mathematics, City University London,\\
$\,\,$ Northampton Square, London EC1V 0HB, UK\\
$^\circ$ Department of Physics, University of Jijel, BP 98, Ouled Aissa,
18000 Jijel, Algeria \\
E-mail: bobphys@gmail.com, a.fring@city.ac.uk}

\input{tcilatex}
\begin{document}

The two dimensional massless Dirac equation has recently attracted a lot of
renewed attention because it describes quasi-particles in graphene \cite%
{graphene1,graphene2,graphene3}, which is well known to possess a large
amount of remarkable properties. Especially the Dirac equation in the
presence of a magnetic field is of great interest as, unlike electrostatic
potentials, such a configuration allows in principle to confine the Dirac
fermions \cite{Alessandro,Alessandro2,Giavaras}. Many exact solutions have
been provided for a variety of time-independent Hamiltonians and magnetic
field configurations \cite{Lukose,GhoshTK,Kuru}, including some for complex
magnetic fields leading to pseudo/quasi-Hermitian interactions \cite%
{PanRoy,HoRoy}. While some solutions for the time-dependent Dirac equation
in 1+1 dimensions have been constructed \cite{Guedes,SouzaD,Mustapha},
little is known about the time-dependent setting with a magnetic field in
2+1 dimensions and no exact solutions have been reported. The aim of this
manuscript is to commence filling that apparent gap. We shall demonstrate
that the Lewis-Riesenfeld method of invariants \cite{Lewis69} is a technique
which can be employed successfully to solve this problem.

We consider here the time-dependent massless Dirac equation in two spacial
dimensions in the form%
\begin{equation}
H\mathbf{\Psi }=i\partial _{t}\mathbf{\Psi },\text{\qquad\ \ \ \ with }%
H(x,y,t)=\mathbf{\sigma }\cdot \mathbf{p},  \label{Dirac}
\end{equation}%
and the two component wave function $\mathbf{\Psi }=(\psi _{1},\psi _{2})$.
The effective Hamiltonian includes $\mathbf{\sigma }=(\sigma _{x},\sigma
_{y})$ comprised of the standard Pauli matrices and $\mathbf{p}$ being the
two-dimensional momentum vector minimally coupled to a vector potential $%
A=(A_{x},A_{y})$ in the standard fashion $\mathbf{p}%
=(a(t)p_{x}+A_{x},b(t)p_{y}+A_{y})$. Besides the time-dependence entering
through the vector potential, we allow here also for explicit time-dependent
factors in front of the momenta, $a(t)$ and $b(t)$. One possibility to think
of these factors is that they result from a time-dependent background, as
discussed in \cite{PhysRevD.90.084005} or alternatively as time-dependent
velocities. The velocity of light, the reduced Planck constant and the
charge are all set to one in our discussion, i.e., $c=\hbar =e=1$.

We make now some specially simply choices for the vector potential by taking 
$A_{x}=0$ and $A_{y}(x,t)=g(t)x$. In addition, we assume that the wave
function separates with a free plane wave moving in the $y$-direction%
\begin{equation}
\mathbf{\Psi }(x,y,t)\mathbf{=}e^{i(ky-\omega t)}\mathbf{\Phi }(x,t),
\label{w1}
\end{equation}%
with wave number $k$ and frequency $\omega $, such that we are left with the
task to solve 
\begin{equation}
\hat{H}\mathbf{\Phi }=i\partial _{t}\mathbf{\Phi },\text{\qquad\ \ \ \ with }%
\hat{H}(x,t)=a(t)\sigma _{x}p_{x}+\left[ kb(t)+g(t)x\right] \sigma
_{y}-\omega \mathbb{I},  \label{phi}
\end{equation}%
for the two component wave function $\mathbf{\Phi }=(\phi _{1},\phi _{2})$.
The first observation we make here is that the form of the explicit
time-dependence does not allow for the standard decoupling of the systems
into a pair of Hamiltonians related to each other by intertwining operators
as common in the time-independent Dirac equation in analogy to standard
supersymmetric quantum mechanics \cite{Witten:1981nf,Cooper}.

We will attempt to solve equation (\ref{phi}) by using the Lewis-Riesenfeld
method \cite{Lewis69} originally designed to solve the time-dependent Schr%
\"{o}dinger equation. The first step in this approach consists of solving
the evolution equation%
\begin{equation}
\frac{dI(t)}{dt}=\partial _{t}I(t)+\frac{1}{i}[I(t),\hat{H}(t)]=0,
\label{Inv}
\end{equation}%
for the Hermitian time-dependent invariant $I(t)$. As usual in this context
we take the invariant to be of the same order and form in the canonical
variables as the Hamiltonian%
\begin{equation}
I(t)=\alpha (t)p_{x}+\beta (t)x+\gamma (t),  \label{inv}
\end{equation}%
where $\alpha (t)$, $\beta (t)$ and $\gamma (t)$ are now unknown
time-dependent matrices. Substituting our Ansatz (\ref{inv}) into the
evolution equation (\ref{Inv}) yields six constraining equations for the
three coefficient matrices%
\begin{equation}
\left[ \alpha ,\sigma _{x}\right] =0,\quad \quad \left[ \beta ,\sigma _{y}%
\right] =0,\quad \quad g\left[ \alpha ,\sigma _{y}\right] +a\left[ \beta
,\sigma _{x}\right] =0,  \label{xxx}
\end{equation}%
\begin{eqnarray}
-i\dot{\alpha} &=&kb\left[ \alpha ,\sigma _{y}\right] +a\left[ \gamma
,\sigma _{x}\right] ,  \label{c1} \\
-i\dot{\beta} &=&g\left[ \gamma ,\sigma _{y}\right] ,  \label{c2} \\
-i\dot{\gamma} &=&kb\left[ \gamma ,\sigma _{y}\right] +ia\sigma _{x}\beta
-ig\alpha \sigma _{y}.  \label{c3}
\end{eqnarray}%
Expanding the matrices in the $su(2)$-basis, $\alpha (t)=\alpha _{1}(t)%
\mathbb{I+}\alpha _{2}(t)\sigma _{x}+\alpha _{3}(t)\sigma _{y}+\alpha
_{4}(t)\sigma _{z}$ with $\alpha _{i}(t)\in \mathbb{R}$ for $i=1,2,3,4$ and
similarly for $\beta (t)$, $\gamma (t)$, these equations are straightforward
to solve. Starting with (\ref{xxx}), the first two equations immediately
imply that $\alpha _{3}=\alpha _{4}=0$ and $\beta _{2}=\beta _{4}=0$. The
last equation in (\ref{xxx}) then yields $\beta _{3}=\alpha _{2}g(t)/a(t)$.
Proceeding in this way for (\ref{c1})-(\ref{c3}), we find the following form
for the time-dependent invariant%
\begin{equation}
I(t)=\left( \alpha _{1}p_{x}+\gamma _{1}\right) \mathbb{I}+\alpha
_{2}p_{x}\sigma _{x}+\left[ \beta _{3}x+\gamma _{3}(t)\right] \sigma _{y},
\end{equation}%
with constants $\alpha _{1},\gamma _{1},\alpha _{2},\beta _{3}$. The
time-dependence of $I(t)$ is entirely contained in the function $\gamma
_{3}(t)$ which is constrained by%
\begin{equation}
\dot{\gamma}_{3}(t)=\alpha _{1}g(t),\qquad \text{and\qquad }\gamma
_{3}(t)=k\alpha _{2}\frac{b(t)}{a(t)}=k\beta _{3}\frac{b(t)}{g(t)}.
\label{gamma}
\end{equation}%
In addition we found that $a(t)=\mu g(t)$ with $\mu =\alpha _{2}/\beta _{3}$
being constant has to be satisfied. The equations (\ref{gamma}) are most
conveniently solved in terms of $b(t)$%
\begin{equation}
\gamma _{3}(t)=\left( 2k\alpha _{1}\beta _{3}\int^{t}b(s)ds\right)
^{1/2},\qquad g(t)=\frac{k\beta _{3}}{2\alpha _{1}}b(s)\left(
\int^{t}b(s)ds\right) ^{-1/2}.
\end{equation}

The next step in the Lewis Riesenfeld approach consists of solving the
eigenvalue equation for the time-dependent invariant, i.e., we need to solve%
\begin{equation}
I(t)\mathbf{\chi }(t)=\lambda \mathbf{\chi }(t),  \label{eigen}
\end{equation}%
for the time-dependent eigenfunction $\mathbf{\chi }(t)=(\chi _{+}(t),\chi
_{-}(t))$ and time-independent eigenvalues $\lambda $. For this purpose we
note at first that we can write (\ref{eigen}) as%
\begin{equation}
\left( 
\begin{array}{ll}
0 & L_{-} \\ 
L_{+} & 0%
\end{array}%
\right) \left( 
\begin{array}{l}
\chi _{+} \\ 
\chi _{-}%
\end{array}%
\right) =\frac{1}{\alpha _{2}}(\lambda -\alpha _{1}-\gamma _{1})\left( 
\begin{array}{l}
\chi _{+} \\ 
\chi _{-}%
\end{array}%
\right) .  \label{LL}
\end{equation}%
Thus we notice that unlike as the time-dependent Hamiltonian the invariant
equation can be decoupled easily and acquires the form of a supersymmetric
pair. Acting again with the off-diagonal invariant operator on (\ref{LL}) we
obtain the two decoupled equations 
\begin{equation}
I_{\pm }\chi _{\pm }=\left[ p_{x}^{2}+W^{2}\pm W^{\prime }\right] \chi _{\pm
}=\frac{1}{\alpha _{2}^{2}}(\lambda -\alpha _{1}-\gamma _{1})^{2}\chi _{\pm
},  \label{III}
\end{equation}%
for the two operators $I_{\pm }:=L_{\mp }L_{\pm }$, where $W(x,t)=[\gamma
_{3}(t)+\beta _{3}x]/\alpha _{2}$ is the analogue to the superpotential in
standard supersymmetric time-independent quantum mechanics. We observe that
the potential is still time-dependent, but now a simple re-definition of our
variables will move this dependence entirely into $\chi _{\pm }$. Defining
the new time-dependent variable $\xi (t)=[\gamma _{3}(t)+\beta _{3}x]/\alpha
_{2}$ converts (\ref{III}) into two eigenvalue equations for the
time-independent quantum harmonic oscillator%
\begin{equation}
\left( -\frac{1}{2}\frac{d^{2}}{d\xi ^{2}}+\frac{\mu ^{2}}{2}\xi ^{2}\right)
\chi _{\pm }=\tilde{\lambda}^{\pm }\chi _{\pm },
\end{equation}%
with $2\tilde{\lambda}_{\pm }=$ $(\lambda -\alpha _{1}-\gamma
_{1})^{2}/\beta _{3}^{2}\mp \mu $. Demanding $\chi _{\pm }(\xi )$ to be a
square integrable function $L^{2}(\mathbb{R},d\xi )$ the solution is of
course%
\begin{equation}
\chi _{\pm ,n}(\xi )=\frac{1}{\sqrt{2^{n}n!}}\left( \frac{\mu }{\pi }\right)
^{1/4}e^{-\mu /2\xi ^{2}}H_{n}(\sqrt{\mu }\xi ),\qquad \tilde{\lambda}%
_{n}^{\pm }=\mu \left( n+\frac{1}{2}\right) ,  \label{w2}
\end{equation}%
with $H_{n}$ denoting the $n$-th Hermite polynomial. This means the two
eigenvalues for the time-dependent spinor components of the invariant in (%
\ref{eigen}) quantize to%
\begin{equation}
\lambda _{n,s}^{\pm }=\alpha _{1}+\gamma _{1}+\beta _{3}s\sqrt{\mu (2n+1\pm
1)},
\end{equation}%
with $s=\pm 1$ being two possible signs of the square root. As expected
these eigenvalues are indeed time-independent. Due to the supersymmetric
structure we have the standard shift in the eigenvalues, that is $\lambda
_{n+1}^{-}=\lambda _{n}^{+}.$

As argued by Lewis and Riesenfeld \cite{Lewis69} the eigenfunction of the
Hamiltonian and the invariant just differ by a phase 
\begin{equation}
\left\vert \mathbf{\Phi }_{n}\right\rangle =e^{i\delta (t)}\left\vert 
\mathbf{\chi }_{n}\right\rangle  \label{ph}
\end{equation}%
where the real function $\delta (t)$ in (\ref{ph}) must obey%
\begin{equation}
\frac{d\delta (t)}{dt}=\left\langle \mathbf{\chi }_{n}\right\vert i\partial
_{t}-\hat{H}(t)\left\vert \mathbf{\chi }_{n}\right\rangle .  \label{pht}
\end{equation}%
The right hand side of (\ref{pht}) can be computed directly with our known
eigenfunctions. We obtain%
\begin{eqnarray}
\left\langle \mathbf{\chi }_{n}\right\vert i\partial _{t}-\hat{H}%
(t)\left\vert \mathbf{\chi }_{n}\right\rangle &=&\left\langle \chi
_{+,n}\right\vert i\partial _{t}+\omega \left\vert \chi _{+,n}\right\rangle
+\left\langle \chi _{+,n}\right\vert -ap_{x}+i(kb+gx)\left\vert \chi
_{-,n}\right\rangle \\
&&+\left\langle \chi _{-,n}\right\vert -ap_{x}-i(kb+gx)\left\vert \chi
_{+,n}\right\rangle +\left\langle \chi _{-,n}\right\vert i\partial
_{t}+\omega \left\vert \chi _{-,n}\right\rangle  \notag \\
&=&2\omega ,
\end{eqnarray}%
where we used the orthogonality of the eigenfunctions, $\left\langle \chi
_{\pm ,n}\right\vert \partial _{t}\left\vert \chi _{\pm ,n}\right\rangle =0$%
, $\left\langle \chi _{+,n}\right\vert p_{x}\left\vert \chi
_{-,n}\right\rangle =-\left\langle \chi _{-,n}\right\vert p_{x}\left\vert
\chi _{+,n}\right\rangle $ and $\left\langle \chi _{+,n}\right\vert
x\left\vert \chi _{-,n}\right\rangle =\left\langle \chi _{-,n}\right\vert
x\left\vert \chi _{+,n}\right\rangle $. The phase therefore simply becomes%
\begin{equation}
\delta (t)=2\omega t.  \label{w3}
\end{equation}%
Thus assembling the results from equations (\ref{w1}), (\ref{w2}), (\ref{ph}%
) and (\ref{w3}) provides an exact solution to the Dirac equation (\ref%
{Dirac}).

We have demonstrated that the Lewis-Riesenfeld method can be applied to
construct solutions to the 2+1 dimensional time-dependent Dirac equation.
The time-dependence resulted from a background and a magnetic field. Exact
solutions for this type of scenario have not been known previously. Clearly
there are plenty of open problems and challenges left. For instance, just as
in the time-independent scenario one would like to know exact solutions for
more complicated vector field configurations, different background scenarios
and possibly different assumptions about the motion in the $y$-direction.
These tasks are left for future work, where this note can be taken as
encouragement as it demonstrates the successful applications of a method to
tackle these kind of problems.

\bigskip \noindent \textbf{Acknowledgments:} BK would like to thank Jijel
University for financial support and City University London for kind
hospitality. AF would like to thank Alessandro de Martino for useful
discussions.

\newif\ifabfull\abfulltrue

\end{document}